\RequirePackage[table,dvipsnames]{xcolor}
\documentclass[journal,twoside,web]{ieeecolor}
\usepackage{generic}
\usepackage{cite}
\usepackage{amsmath,amssymb,amsfonts}
\usepackage{algorithmic}
\usepackage{graphicx}
\usepackage{algorithm,algorithmic}
\usepackage{xr-hyper}
\usepackage{hyperref}
\usepackage{xurl}
\usepackage{pdfpages}
\input{supplabels}
\usepackage{booktabs}
\usepackage{tabularx}
\usepackage{threeparttable}

\hypersetup{hidelinks}
\usepackage{textcomp}
\usepackage[strings]{underscore}
\usepackage[utf8]{inputenc}
\usepackage{newunicodechar}
\newunicodechar{↔}{\leftrightarrow}

\usepackage{rotfloat}
\usepackage[acronym]{glossaries}
\glsdisablehyper
\usepackage{array}
\usepackage{float}
\usepackage{tikz}
\graphicspath{{figures/}}

\usetikzlibrary{shapes.geometric, arrows.meta, positioning, fit, backgrounds}

\definecolor{ethblue}{RGB}{33,92,175}
\definecolor{ethdark}{RGB}{20,55,105}
\definecolor{ethlight}{RGB}{210,225,245}
\definecolor{ethgray}{RGB}{80,80,80}
\definecolor{storagegray}{RGB}{245,245,250}

\newacronym{AFH}{AFH}{adaptive frequency hopping}
\newacronym{API}{API}{application programming interface}
\newacronym{BLE}{BLE}{Bluetooth Low Energy}
\newacronym{EMC}{EMC}{electromagnetic compatibility}
\newacronym{FDA}{FDA}{Food and Drug Administration}
\newacronym{GATT}{GATT}{Generic Attribute Profile}
\newacronym{HCI}{HCI}{Host Controller Interface}
\newacronym{ISM}{ISM}{industrial, scientific, and medical}
\newacronym{JSON}{JSON}{JavaScript Object Notation}
\newacronym{LSL}{LSL}{Lab Streaming Layer}
\newacronym{RMS}{RMS}{root mean square}
\newacronym{SPZ}{SPZ}{Swiss Paraplegic Center}
\newacronym{CPU}{CPU}{central processing unit}
\newacronym{QBSS}{QBSS}{QoS Basic Service Set}
\newacronym{PMD}{PMD}{Polar Measurement Data}
\newacronym{RSSI}{RSSI}{received signal strength indicator}
\newacronym{REST}{REST}{Representational State Transfer}
\newacronym{UDP}{UDP}{User Datagram Protocol}
\newacronym{UUID}{UUID}{universally unique identifier}
\newacronym{bcg}{BCG}{ballistocardiogram}
\newacronym{ecg}{ECG}{electrocardiogram}
\newacronym{eda}{EDA}{electrodermal activity}
\newacronym{eeg}{EEG}{electroencephalogram}
\newacronym{emg}{EMG}{electromyogram}
\newacronym{eog}{EOG}{electrooculogram}
\newacronym{icg}{ICG}{impedance cardiogram}
\newacronym{imu}{IMU}{inertial measurement unit}
\newacronym{ppg}{PPG}{photoplethysmogram}
\newacronym{ros2}{ROS~2}{Robot Operating System~2}
\newacronym{wban}{WBAN}{wireless body area network}
\newacronym{QoS}{QoS}{Quality of Service}
\newacronym{crc}{CRC}{Cyclic Redundancy Check}
\newacronym{HTTPS}{HTTPS}{Hypertext Transfer Protocol Secure}
\newacronym{MAC}{MAC}{Medium Access Control}
\newacronym{CI}{CI}{confidence interval}
\begin{document}

\title{Connected but Late: Sample Age of Commercial BLE Wearables under Wi-Fi Coexistence for Real-Time Clinical Applications}
\author{Monica Perez-Serrano, H\"umay Zehra \"Ozer, Seung Ju Paek, and Diego Paez-Granados
\\[3pt] {\normalfont\scriptsize\itshape This work has been submitted to the IEEE for possible publication. Copyright may be transferred without notice, after which this version may no longer be accessible.}\vspace{-14pt}
\thanks{This work was supported in part by the Open Research Data program of the ETH Board, and in part by the Schweizer Paraplegiker Stiftung and ETH Z\"urich under Grant 2021-HS-348 within the Digital Transformation in Personalized Healthcare in SCI and SPS-FOKO. This work involved human subjects in its research. Approval was granted by the Ethikkommission Nordwest- und Zentralschweiz (EKNZ) under Application No. 2025-00165.
\textit{(Corresponding authors: Monica Perez-Serrano and Diego Paez-Granados.)}}
\thanks{M. Perez-Serrano is with the SCAI Lab, D-HEST, ETH Z\"urich, 8092 Zurich, Switzerland, the Lake Lucerne Institute, 6354 Vitznau, Switzerland, and the Swiss Paraplegic Research (SPF), 6207 Nottwil, Switzerland (e-mail: monica.perezserrano@hest.ethz.ch).}
\thanks{H. Z. \"Ozer is with the SCAI Lab, D-HEST, ETH Z\"urich, 8092 Zurich, Switzerland, and the University of Bern, 3012 Bern, Switzerland.}
\thanks{S. J. Paek is with the University of Zurich (UZH), 8006 Zurich, Switzerland.}
\thanks{D. Paez-Granados is with the SCAI Lab, D-HEST, ETH Z\"urich, 8092 Zurich, Switzerland, and the Swiss Paraplegic Research (SPF), 6207 Nottwil, Switzerland, and the Cybernics Institute at University of Tsukuba, Japan (e-mail: diego.paez@hest.ethz.ch).}}

\maketitle

\begin{abstract}
Wearable sensors are increasingly being used in real-time clinical applications, such as rehabilitation robots and live therapy monitoring. As Healthcare~5.0 moves towards personalized therapies, their use will continue growing, potentially compromising communication reliability. Most wearables transmit over Bluetooth Low Energy, sharing 2.4\,GHz with institutional Wi-Fi. Previous coexistence studies have focused on whether data arrive, rather than when. We measured sample age for a commercial electrocardiogram strap, from acquisition through the gateway to storage in our platform MEDWEAR, in an anechoic chamber and a hospital rehabilitation gym, and we compared it with illustrative clinical deadlines. Without load, the strap's own framing was the largest source of delay, making samples on average 280\,ms and up to 550\,ms old before transmission. Wi-Fi load had no measurable effect while adaptive frequency hopping could avoid the loaded channel. In a stress test, which forced the link onto the loaded channel, the link became the dominant source of delay, adding up to 11\,s. The link stayed connected, but a median of 29\% of samples and up to 98\% (worst case) were older than 1\,s on arrival at the gateway. The gym behaved mostly like the quiet chamber, but during one traffic burst, while hopping was restricted, up to 4.6\% of samples were older than 1\,s. These results show that a connected link can deliver late data. Therefore, sensors intended for real-time clinical applications should be profiled in situ and monitored based on sample age rather than connection status alone.
\end{abstract}


\begin{figure*}[t]
  \centering
  \includegraphics[width=\textwidth]{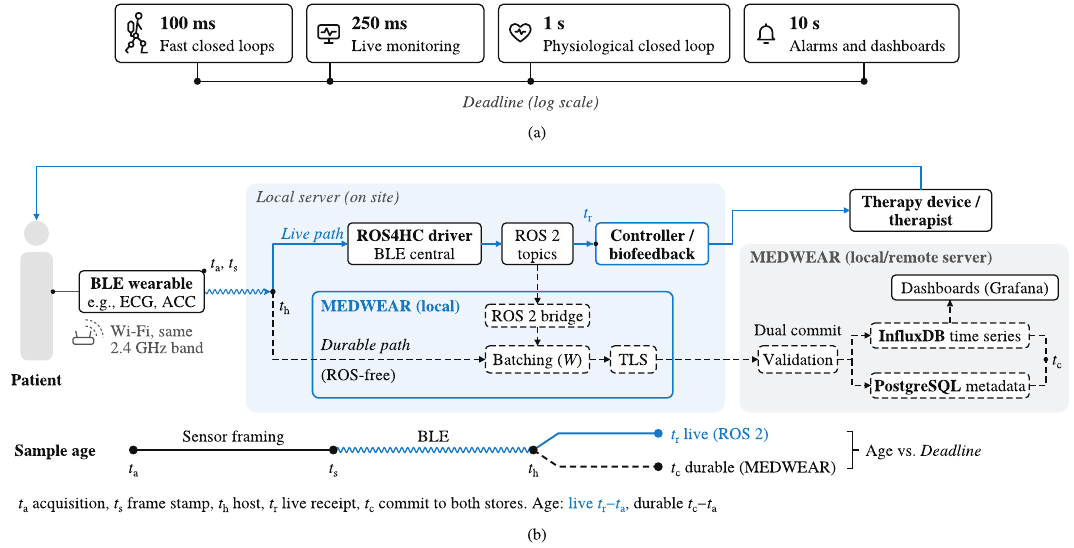}
  \caption{(a) Deadlines of real-time clinical applications, from fast closed loops (100 ms) to alarms and dashboards (10 s). (b) BLE wearable data, sharing the 2.4 GHz band with Wi-Fi, follow a live ROS 2 path to the controller and therapy device, and a durable MEDWEAR path to storage and dashboards. Sample age is then compared with each deadline.}
  \label{fig:medwear-pipeline}
\end{figure*}

\section{Introduction}
\label{sec:introduction}

\IEEEPARstart{W}{earable} health technologies are shaping healthcare by enabling continuous monitoring beyond clinical settings, and are becoming key to personalized and preventive care \cite{babu_wearable_2024,hameed_integrating_2024,tyler_real-time_2020}. Industry~5.0 pairs humans with machines \cite{maddikunta_industry_2022}, and in Healthcare~5.0, these devices are not passive loggers but nodes in a \gls{wban} that feed systems acting on what they measure \cite{younesi_healthcare_2026}. Rehabilitation robotics is the clearest instance, where biosignals must reach a controller within a bounded latency for the loop to close \cite{manzoori_adaptive_2025}.

Most commercial wearables stream over \gls{BLE}, which divides the 2.4\,GHz band into 40 channels of 2\,MHz, three for advertising and 37 for data \cite[Vol.~6, Part~A, Sec.~2, and Part~B, Sec.~1.4.1]{bluetooth_core_54}. Because \gls{BLE} and Wi-Fi share this unlicensed band, \gls{BLE} copes with interference through \gls{AFH}, hopping only over the data channels that the central classifies as ``good'' \cite[Vol.~6, Part~B, Sec.~4.5.8]{bluetooth_core_54}. In hospitals, Wi-Fi concentrates on channels 1, 6, and 11 \cite{al_kalaa_long_2016,al_kalaa_characterizing_2017}, which leaves only 5 of the 37 data channels entirely outside their nominal 20\,MHz bands, and 9 outside their 16.25\,MHz occupied bands. \gls{AFH} protects a link only if the central classifies channels correctly, which can fail under interference \cite{spork_improving_2020}. 

Hospital surveys showed the 2.4\,GHz band to be largely idle, as its utilization stayed below 10\% in 97-99\% of observations, but with sporadic peaks above 50\% \cite{al_kalaa_long_2016,al_kalaa_characterizing_2017}. To handle this risk, U.S.\ \gls{FDA} guidance calls for coexistence testing \cite{health_radio_2020}, for which the \gls{FDA} recognizes ANSI C63.27 \cite{c63.27_ieee_2022}, with AAMI TIR69 \cite{aami_tir69_2017} for risk management. C63.27 checks that a device maintains a manufacturer-specified functional wireless performance. However, passing such a test does not guarantee usable data. In chamber evaluations, a \gls{BLE} link kept delivering data under Wi-Fi on channels 1, 6, and 11, but its throughput fell from 340 to 100\,kbit/s \cite{al_kalaa_case_2020}. Clinical deployments showed the same gap. Wireless sensors delivered 36-98\% of vital-sign data on hospital wards, with connectivity among the causes \cite{wilson_completeness_2025}, and 39-87\% across ward and home \cite{jonkers_data_2026}.  

Among mitigation strategies, lowering the rate, bundling samples, and re-requesting lost packets reduced packet loss below 1\% \cite{tipparaju_mitigation_2021}. At the link layer, studies have quantified packet error, throughput, and latency under interference \cite{karvonen_interference_2017,la_dense_2018,pang_bluetooth_2022}, and new hopping and scheduling schemes have improved timeliness, but only on programmable radios \cite{poirot_eafh_2022,li_rtble_2023}. In simulation, an adaptive \gls{wban} \gls{MAC} protocol kept the packet delay of medical traffic below the 125\,ms bound of IEEE 802.15.6 \cite{hassan_adaptive_2024}.

Spörk et al.\ \cite{spork_improving_2020} measured the share of \gls{BLE} packets exceeding a delay bound under Wi-Fi, from custom-firmware boards to commercial centrals. They showed that \gls{AFH} fails when the interference is near the peripheral, and adapted the connection interval from \gls{HCI} timing. With modifiable firmware, samples can be timestamped to the receiver with 47\,$\mu$s \gls{RMS} jitter \cite{biagetti_high-accuracy_2025}.

Aside from the radio link, real-time frameworks such as \gls{LSL} \cite{kothe_lab_2025}, BRAND \cite{ali_brand_2024}, and \gls{ros2} \cite{macenski_ros2_2022} synchronize and fuse biosignals with low latency, and wearable data platforms such as RADAR-base \cite{ranjan_radar-base_2019} and mCerebrum \cite{hossain_mcerebrum_2017} are mainly evaluated on how much data they collect, not on how old they are.

In practice, developers build on commercial wearables whose firmware they cannot modify, and whether a deadline is met depends on the sensor, the gateway, the pipeline, and the environment. We characterize timeliness as the ability to meet a specified latency budget reliably, inspired by 5G ultra-reliable low-latency communication (URLLC) but without its performance requirements \cite{etsi_tr138913_v17,bennis_ultrareliable_2018}. We follow biosignal streams through sensor acquisition, \gls{BLE} link, queueing, and the host pipeline, from an anechoic chamber through a rehabilitation gym into MEDWEAR, our open-source platform.

This work makes two contributions. First, we propose a profiling approach for commercial wearables whose firmware cannot be modified. For sensors that timestamp their frames, it separates the sample age set by the sensor's framing from the delay added by the link, and compares it with illustrative clinical deadlines from 100\,ms to 10\,s. We apply it in a rehabilitation gym and, extended into a stress test inspired by ANSI C63.27 \cite{c63.27_ieee_2022}, in an anechoic chamber, to an \gls{ecg} strap and a wrist device. Second, we present MEDWEAR, an open platform that implements the profiling tool and carries sensor timestamps to the consumer through standardized data schemas. It offers a live \gls{ros2} path and a durable path, evaluated by replaying the chamber and gym recordings on a single node and by ingesting buffered data from a clinical study.

\section{Methods}
\label{sec:methods}
\subsection{Devices and Drivers}
\label{sec:devices}

We characterize two devices on their native \gls{BLE} \gls{GATT} links: the consumer \gls{ecg} chest strap Polar H10 (firmware 5.0.0) and the medical-grade wrist device Corsano CardioWatch 287-2B (firmware 8.65). Corsano is a CE-marked Class~IIa medical device (EU Medical Device Regulation 2017/745, \gls{BLE}~5.0, 0\,dBm) \cite{corsano_287_2} that buffers up to 24\,h of data in flash \cite{corsano_ifu}. Its \gls{EMC} certification under IEC~60601-1-2 tests immunity, but not coexistence \cite{corsano_ifu,al_kalaa_case_2020}. Moreover, the two links transfer data differently. The Polar H10 pushes data via its \gls{PMD} service: it streams \gls{ecg} at 130\,Hz in packets of 73 samples (one notification every 0.56\,s) and accelerometer data at a nominal 200\,Hz (203.9\,Hz on the strap's clock, $\pm8$\,g, 16\,bit) in packets of 36 samples. When heart rate is detected, it sends an additional notification per second, giving 8.45 notifications/s in total. Instead, the Corsano's data are pulled: the watch writes measurements to files, and the driver repeatedly requests the newest bytes of such files, which the watch returns in 2.7 notifications per request. For this study, the driver is configured to request bio-impedance, accelerometer, heart rate, and stress data. This is not the Corsano's intended use, as it nominally transmits data only about once per minute \cite{corsano_287_2}. Both drivers came from ROS4HC \cite{manriquez-cisterna_ros_2026}. The Polar H10 link used a 45\,ms connection interval and 6\,s supervision timeout, the Corsano link 60\,ms and 5\,s, both on the 2M physical layer (PHY) and unchanged after connection setup. Although placed at the same location, at no load, the Corsano's signal was 29\,dB weaker ($-74$ vs.\ $-45$\,dBm, Supplementary Table~\ref{S-tab:s-link}).

To keep the sensors streaming, both wearables were strapped around a saline-moistened sponge, with electrode gel added on the Polar H10 electrodes, and a wet tissue with conductive paste under the Corsano sensor. The signals are not physiological, and body shadowing is not modeled. The \gls{BLE} central was a laptop with an Intel AX211 controller running BlueZ (versions in Supplementary Section~\ref{S-sup:instruments}). Its internal Bluetooth and Wi-Fi modules share the antenna, so the Wi-Fi module was soft-blocked in every run. 

\subsection{MEDWEAR Platform}
\label{sec:platform}
MEDWEAR is an open, self-hostable platform for raw wearable biosignals \cite{medwear_software_2026} (Fig.~\ref{fig:medwear-pipeline}). It uses standardized raw-biosignal \gls{JSON} schemas and their \gls{ros2} counterparts \cite{perez_serrano_healthcare_msgs}, which carry sensor timestamps end to end (further details in Supplementary Section~\ref{S-sup:platform}).

\subsubsection{Meeting Clinical Deadlines}
Depending on the target application, the consumer will have a different latency budget (Table~\ref{tab:deadlines}), which will determine which path to follow. The \emph{live path} is intended for up to 1\,s deadlines, where data are published as \gls{ros2} topics (using the ROS4HC drivers) on the gateway and consumed directly. The \emph{durable path} is instead for dashboards, alarms, and analysis, where the data are batched over a window $W$ and sent in the standardized \gls{JSON} format to the ingestion service via \gls{HTTPS}. Through \texttt{ros2_to_medwear}, live-path data can also be sent over the durable path, for instance to keep a copy in MEDWEAR. Alternatively, vendor-specific files can also be ingested offline through device-specific parsers. The ingestion service validates each batch (structure, timestamp order, and sampling rate when given), stores a checksum, writes metadata to PostgreSQL and samples to InfluxDB. The acknowledgment only comes once all sent data have been successfully ingested. An idempotency key answers a resubmitted complete batch as a duplicate and completes a pending one.

\subsubsection{Environment Profiling}
\label{sec:profiling-tool}

\begin{figure}[t]
  \centering
  \includegraphics[width=\columnwidth]{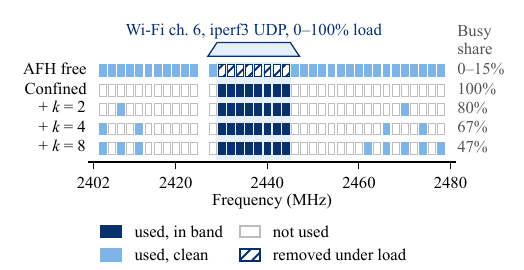}
  \caption{Channel sets in the chamber. With \gls{AFH} free, the controller chose the hop set and, under load, removed the channels overlapped by Wi-Fi channel~6 (hatched). Confined runs held the link on data channels 12-19 inside that band, alone or with $k$ clean channels outside it. Busy share: fraction of the hop set inside the band (\gls{AFH} free: run means under load).}
  \label{fig:confinement}
\end{figure}

MEDWEAR's profiling tool characterizes a sensor's link where it will be used, from the gateway with a Wi-Fi adapter in monitor mode and a \gls{BLE} sniffer. A profiling session proceeds as follows:
\begin{enumerate}
  \item A passive scan of Wi-Fi channels 1-13 identifies the busiest channel.
  \item Runs alternate between free and confined \gls{AFH} (Fig. \ref{fig:confinement}). Three runs of 300\,s per condition, with $k\in\{0,2,8\}$, in a random order whose seed is logged. The host limits the possible hopping channels to the \gls{BLE} ones inside the busiest Wi-Fi channel's (step 1) band plus the $k$ quietest channels outside it.
\end{enumerate}
During each run, the tool records the \gls{HCI} traffic of the devices, and the sensor data with their timestamps. From the monitor adapter, it records the airtime of decoded frames (a lower bound) and the number of beaconing networks per Wi-Fi channel at 15\,s intervals. To protect the privacy of individuals present during recording, raw Wi-Fi monitor-mode captures are not permanently retained, as they may reveal third-party devices and networks through \gls{MAC} addresses and network names. Instead, only per-channel airtime measurements and beacon counts are stored. Deployment of this tool is, however, subject to site approval.

\subsection{Anechoic Chamber}
\label{sec:chamber}

\begin{figure}[t]
  \centering
\includegraphics[width=\linewidth]{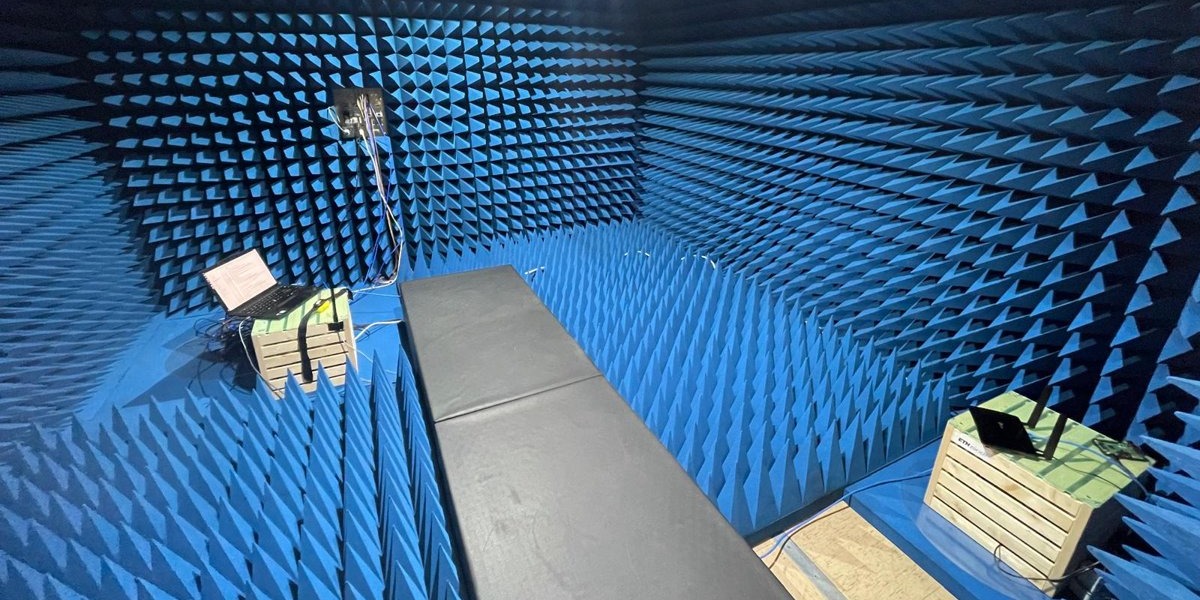}
  \caption{Anechoic chamber: platform~A (left: wearable, central, \gls{BLE} sniffer, Wi-Fi monitor) and platform~B (right: access point and traffic source), 2.28\,m apart on the absorber-covered floor. Traffic ran from the source over Wi-Fi to the access point and over Ethernet to the host.}
  \label{fig:chamber-setup}
\end{figure}

The measurements took place in a radio-frequency shielded room (Ray Proof Corp., $3.7\times3.7\times2.3$\,m) lined with pyramidal absorbers at the Integrated Systems Laboratory, ETH Z\"urich (Fig.~\ref{fig:chamber-setup}), on two wooden platforms with an acrylic top. An ambient scan showed no foreign Wi-Fi network. Platform A carried the wearable (one at a time), the central (laptop, 10\,cm behind the wearable), an nRF52840 sniffer, and the Wi-Fi monitor. At 2.28\,m, Platform B had a TP-Link TL-MR6400 access point on channel~6 (2437\,MHz, 20\,MHz, band steering and \gls{QoS} disabled) and a Raspberry Pi~3B+. Wi-Fi load was induced by the Raspberry Pi via \texttt{iperf3} \gls{UDP} traffic (1200-byte payload, not adapting to loss) at 25-100\% of the \gls{UDP} saturation throughput measured at the start of each session (about 62\,Mbit/s in both sessions). Because \gls{BLE} collisions depend on airtime, the monitor read the access point's \gls{QBSS} Load channel utilization every 30\,s.

Each run recorded the data specified in Section~\ref{sec:profiling-tool} (\texttt{btmon}, \gls{ros2} bag), and lasted 300\,s from connection (1800\,s for the long runs). After device discovery and 30\,s of settling, load began 41–48\,s after the capture started, resulting in 265–272\,s of loaded data per run (approximately 1770\,s in the long runs). Run order was randomized within each block using a logged seed. Conditions (Fig.~\ref{fig:confinement}) were:

\begin{itemize}
  \item \emph{\gls{AFH} free:} 0–100\% load in 25-point increments, three replicates per level, plus a no-load bracket. Additionally, two longer runs of 1800\,s only at 0 and 100\%.
  \item \emph{Confined:} channels 12–19, inside Wi-Fi channel ~6. 0–100\% load, three replicates per level, plus a no-load bracket.
  \item \emph{Confined plus $k$ clean channels:} $k \in \{0,2,4,8\}$ channels selected from 0–5 and 27–36: $\{2,32\}$ for $k=2$, $\{0,4,30,34\}$ for $k=4$, and $\{0,2,4,28,30,32,34,36\}$ for $k=8$. Each configuration had three replicates at 75\% load, with additional no-load runs at $k=2$ and 100\%-load runs at $k\in\{2,8\}$ (for the Polar H10, in session~2).
\end{itemize}

The channel map was applied after connection setup and before load onset, and it held through the Corsano's reconnections: all channel-map samples of its 15 loaded confined runs showed 8 channels in use. A second Polar H10 session was added after the first had been analyzed, because the first session showed a more robust link than an uncontrolled bench pre-test, run with the central's own Wi-Fi on. The second session interleaved two confined and two $k=0$ runs at 75\% load, recorded $k\in\{2,8\}$ at 100\%, and added two confined no-load runs. Both labels used the same channel map, so the confined and $k=0$ runs at 75\% load are pooled as the $k=0$ condition (ten Polar H10 runs).

\subsubsection*{What confinement emulates} 
A forced map retains busy channels, as may occur when a central's classification fails or the rest of the band is busier. Adding $k$ clean channels reduces the realized busy share to 100, 80, 67, and 47\% for $k=0,2,4,8$, respectively (Fig. \ref{fig:confinement}). This reproduces the hop set of such a band, not its radio environment. A central unable to classify channels would retain all 37 channels, of which 22\% are busy. Thus, $k=0$ is a stress test beyond observed states (Section~\ref{sec:results-chamber}). Before each \gls{AFH}-free run, the host classification was reset to all 37 channels, whereas the controller's classification persisted.

\subsection{Gym Recordings}
\label{sec:gym-methods}
The gym recordings used MEDWEAR's profiling tool (Section~\ref{sec:profiling-tool}) in the rehabilitation gym of the \gls{SPZ} during regular therapy hours, in three sessions on two days (13 and 16 October 2026) covering high and low occupancy. The setup was as in the chamber, at ${\sim}1.5$\,m height, and no artificial Wi-Fi load was generated. Confined runs used $k\in\{0,2,8\}$. The gym's $k=8$ set (10, 22, and the even channels 26-36) included four channels under Wi-Fi channel~11, so it is not strictly comparable to the chamber's. Each condition ran 300\,s, three replicates, interleaved, and a head count was logged manually about every 30\,min. Without our own access point, airtime came from decoded frames in monitor mode, not \gls{QBSS} Load. No personal or health data were collected (Section~\ref{sec:profiling-tool}), so the recordings fall outside the Swiss Human Research Act. The \gls{SPZ} permitted recording during regular operation.

\subsection{Trace-Driven Replay through MEDWEAR}
\label{sec:replay}

The chamber and gym recordings were replayed through MEDWEAR with their original receive timing (\texttt{ros2 bag play}, rate 1.0), so delays, bursts, and gaps reached the platform as received. The bridge closed a batch when its oldest sample had waited $W$, checked at each message and every 50\,ms, or early at gaps over 1\,s. All streams were replayed at $W=1$\,s, and the Polar H10 \gls{ecg} of Section~\ref{sec:results-e2e} also at $W\in\{0.1,0.5,5\}$\,s. Replay and MEDWEAR (including the \gls{ros2} bridge and the production limit of 300 batch requests per minute) ran together in Slurm jobs with exclusive allocation of one node (AMD EPYC 9555P, 64 cores, rootless Podman 5.8.2, \gls{CPU} time sampled at 1\,Hz), so all replay times share one clock, whose rate error (about 60\,ppm) is negligible over the sub-second intervals measured. The bridge recorded when it received each sample and sent each batch, and the \gls{API} when it received, validated, and committed each batch. Each sample's commit is added to the original timeline as $t_\text{c}=t_\text{r}+(\text{commit}-\text{bridge receipt})$, both from the replay. 

\subsection{Ingestion of Buffered Clinical Data}
\label{sec:scale-methods}
To test the ingestion of buffered offline data, we ingested, through a custom offline parser, the vendor exports of the CARDAMON study (Swiss Paraplegic Center, Ethikkommission Nordwest- und Zentralschweiz, 2025-00165, with informed consent from all participants). In this configuration, Corsano recordings held \gls{eda}, and an additional sensor (VivaLink) provided \gls{ecg}. Finally, we also tested MEDWEAR under load and faults on one node (Supplementary Section~\ref{S-sup:benchmark}).

\subsection{Metrics}
\label{sec:metrics}
\subsubsection*{What each device allows} The Polar H10 stamps every frame with sensor time, so each \gls{ecg} sample's age on arrival is known. The Corsano writes its data to files without sample times, so for it we measure only how much data arrives and how long the host waits for new data. Link-layer measures come from a passive \gls{BLE} sniffer, which did not decrypt the Corsano's link (an LE Secure Connections bond), so they are available only for the Polar H10.

\subsubsection*{Time points} Four time points are defined for each sample:
\begin{itemize}
  \item $t_\text{a}$: acquisition of a sample in the strap.
  \item $t_\text{h}$: arrival at the host of the frame's last fragment (\gls{HCI} receive time), the end of the radio link.
  \item $t_\text{r}$: receipt by a \gls{ros2} subscriber (driver) on the live path.
  \item $t_\text{c}$: commit to both stores on the durable path.
\end{itemize}

\subsubsection*{Sample age (Polar H10)} The strap batches 73 \gls{ecg} samples (0.56\,s) into a packet and sends it with the last sample's timestamp, $t_\text{s}$. So a sample's age at the host has two parts:
\begin{equation}
t_\text{h}-t_\text{a} = \underbrace{t_\text{s}-t_\text{a}}_{\text{framing}} + \underbrace{t_\text{h}-t_\text{s}}_{\text{added delay}},
\label{eq:age}
\end{equation}
where framing is set by the sensor (sample $j=0,\dots,72$ is $(72-j)/130$\,s older than the timestamp, up to 0.55\,s) and added delay is what the link adds: queueing in the strap, radio transmission, and host reception. As the strap's clock is not synchronized with the host's, added delay is measured from its smallest value in the run, so ages are a lower bound (Section~\ref{sec:limitations}). Using the run's 1st percentile instead changed $M'$ (defined below) by at most 6.2, 3.1, and 0.9 percentage points at 100\,ms, 250\,ms, and 1\,s and changed no runaway classification (Supplementary Section~\ref{S-sup:sensitivity}). Clock drift (220 and $-7$\,ppm in the chamber sessions, mostly the host's) is removed with one rate per session, the median lower-envelope slope of added delay in no-load runs (in the gym, unconfined runs). Doing it per run would instead hide a growing backlog. The error is at most 6\,ms in a 300\,s run (Supplementary Section~\ref{S-sup:instruments}). On the live and durable paths, age is defined as $t_\text{r}-t_\text{a}$ and $t_\text{c}-t_\text{a}$.
 
\subsubsection*{Deadline misses (Polar H10)} Each application in Table~\ref{tab:deadlines} needs samples younger than a deadline $D$, which is an illustrative budget, rather than a validated clinical requirement. The deadline-miss ratio $M(D)$ is the share of samples older than $D$ on arrival. Samples that never arrived count as missed, but samples acquired within $D$ of the end of the run are not scored. At 10\,s, this excludes most of an end-of-run backlog, which the durable path counts. $1-M(D)$ is thus the reliability within latency budget $D$. $M'(D)$ repeats this count on added delay alone, i.e. excluding framing. 

\subsubsection*{Data received and staleness (Corsano)} \emph{Data received} is the payload in bytes per second relative to the median without load. \emph{Staleness} is the time since new bio-impedance bytes last reached the host, sampled every 0.1\,s, which is about 2\,s without load. As the files carry no sample times, staleness is a lower bound on the age of information \cite{kaul_aoi_2012,yates_aoi_survey_2021}.

\subsubsection*{Link measures} For both devices, a \emph{link loss} is a disconnection that the host did not initiate (any reason other than 0x16), from 15\,s after the capture started until the driver stopped. Failed reconnection attempts (0x3E) are counted separately. The \emph{busy share} is the fraction of the channels in use that lie inside the busy Wi-Fi channel's band, from the channel map read every 15\,s. For the Polar H10, the sniffer gives \gls{crc} failures, \textit{More-Data} bits, and new packets per connection event, but only for the events it followed. In the same interference, it heard nearly all events on clean channels but only 27-31\% on busy ones in session~1 and 21-22\% at $k=0$ in session~2.

\subsubsection*{Statistics} The run is the independent unit. With three runs per condition in general, and ten at $k=0$ and 75\% load, medians and ranges over runs are reported. Counts of runs carry Wilson 95\% \glspl{CI}, counts of zero carry one-sided exact (Clopper-Pearson) 95\% upper bounds, and proportions of runs are compared with Fisher's exact test. A run fails a deadline when more than 1\% of its samples miss it, a threshold set after the experiments. Because the blocks of session~1 ran in sequence, condition is partly confounded with run order (for example, sponge drying, battery level, or controller state), so runaways are reported by block and session (Supplementary Table~\ref{S-tab:s-zeroheadroom}).

\begin{table}[t]
\centering
\caption{Deadlines at Which $M(D)$ Is Read, and Their Basis}
\label{tab:deadlines}
\scriptsize
\setlength{\tabcolsep}{3pt}
\begin{tabular}{@{}>{\raggedright\arraybackslash}p{0.7cm}>{\raggedright\arraybackslash}p{2.0cm}>{\raggedright\arraybackslash}p{5.5cm}@{}}
\toprule
$D$ & \textbf{Consumer} & \textbf{Basis} \\
\midrule
100\,ms & Fast loops acting on movement & Exoskeleton assistance helped balance at ${\sim}68$\,ms, not ${\sim}171$\,ms \cite{beck_exo_2023}. Prosthesis delay optimum 100-125\,ms \cite{farrell_delay_2007}. 85 vs.\ 25\,ms feedback latency cut implicit learning by a third \cite{hadjiosif_latency_2025}. \\
250\,ms & Live monitoring (therapist following the session) & End-to-end requirement for in-hospital remote monitoring and wearables \cite{cisotto_requirements_2020}. \\
1\,s & Physiological loops (heart rate control) & Our choice: one fifth of a 5\,s heart rate controller update \cite{wang_hrcontrol_2023}, small against the control period. \\
10\,s & Alarms and dashboards & \gls{ecg} monitors must alarm on cardiac standstill within 10\,s \cite{aami_iec_60601_2_27_2016} (upper bound) \\
\bottomrule
\end{tabular}
\par\smallskip
\parbox{\linewidth}{\scriptsize Engineering reference values for the whole chain from acquisition to the consumer, applied to the age of each sample. We read all four on the \gls{ecg} stream. A movement loop would use the accelerometer, whose 0.18\,s frames alone make 42\% of samples older than 100\,ms (57-58\% on arrival with \gls{AFH} free, Supplementary Table~\ref{S-tab:s-sensitivity}), and a clinical alarm requires a regulated device.}
\end{table}

\section{Results}
\label{sec:results}
\subsection{Anechoic Chamber}
\label{sec:results-chamber}
Table~\ref{tab:headroom} summarizes the chamber by hop set, and Fig.~\ref{fig:deadline} shows the Polar H10 at 75\% load. Per-condition and per-run values are in Supplementary Section~\ref{S-sup:chamber}.
 
\subsubsection*{Framing sets the floor}
Before any radio delay, waiting for a 73-sample frame makes 81\% of Polar H10 samples older than 100\,ms and 55\% older than 250\,ms (Table~\ref{tab:headroom}). A freely hopping link added little to this: 63\% were older than 250\,ms and at most 0.1\% older than 1\,s, with or without load. The added delay exceeded 100\,ms for only 2-4\% of samples, a floor set by the 45\,ms connection interval rather than by Wi-Fi.
 
\subsubsection*{With AFH free, Wi-Fi load had no effect}
Neither device lost its link in 18 runs each at 0-100\% load, and sample age did not change. The controller kept the link off the loaded channel and its classification carried over between runs. So busy channels made up at most 15\% of the Polar H10's hop set under load.
 
\subsubsection*{Forced onto the loaded channel, the link stayed up but data aged}
With every hop on the busy channels, the Polar H10 never lost its link (19 loaded runs). At 25, 50, and 100\% load, samples aged somewhat, but at most 0.9\% were older than 1\,s. At 75\% load, more than 1\% of samples were older than 1\,s in eight of ten runs (Wilson 95\% CI 49-94\%), and four ran away (Fig.~\ref{fig:deadline}b): the link sustained only 92-97\% of the strap's notification rate, so a backlog built in the strap's transmit queue, peaked at 9-11\,s, and did not clear, leaving 83-98\% of samples older than 1\,s. One run built a shorter backlog (2.4\,s), and five stayed bounded. Runaways occurred under both labels and in both sessions, Supplementary Section~\ref{S-sup:sensitivity} gives the packet-level mechanism. \gls{ecg} samples were delayed, not lost, but once a backlog exceeded about 6\,s, the strap discarded whole accelerometer frames (0.3-8.4\% of samples).
 
\subsubsection*{Two clean channels prevented the runaway}
\begin{figure*}[t]
  \centering
  \includegraphics[width=0.83\textwidth]{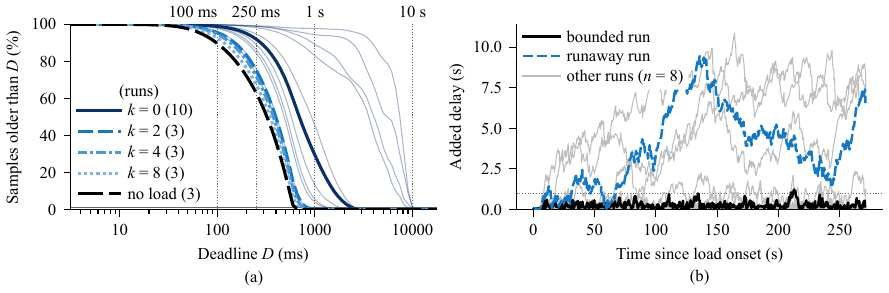}
  \caption{Polar H10 \gls{ecg} timing in the chamber at 75\% Wi-Fi load. (a) Share of samples older than a deadline, $M(D)$, median over runs: all ten $k=0$ runs, three at each other $k$, and three without load (thick lines). Thin lines, single $k=0$ runs. Gray, 1\%. Dotted, deadlines of Table~\ref{tab:deadlines}. (b) Added delay over time in the ten $k=0$ runs: one bounded and one runaway run, both from session~2, and eight others (gray). Dotted line, 1\,s.}
  \label{fig:deadline}
\end{figure*}

With $k\geq2$, no run ran away and neither device lost its link (0/15 loaded runs each, upper 95\% bound 18.1\%). At most 1.2\% of Polar H10 samples were older than 1\,s (Fig.~\ref{fig:deadline}a), and only one run exceeded 1\%, against eight of ten at $k=0$ (Fisher $p=0.0003$, post hoc, or $p=0.005$ against the nine runs at 75\% load) and runaways in 4 of 10 ($p=0.017$). Each added clean channel lowered the share older than 250\,ms (Table~\ref{tab:headroom}), as packets failed mainly on busy channels (\gls{crc} failures 23-43\% per run vs.\ $\leq$0.02\% on clean ones).
 
\subsubsection*{The wrist device disconnected instead}
Without a clean channel, the Corsano lost its link in all 15 loaded runs. Every first loss was a supervision timeout, which depends on whether any packet gets through rather than on the transfer model, so its 29\,dB weaker link or shorter supervision timeout, not polling alone, could explain it. With $k\geq2$ it stayed connected but received only 51-87\% of its data at 75\% load (Table~\ref{tab:headroom}), because each request waits for its reply and requests slowed.

\begin{table}[t]
\caption{Chamber Results by Hop Set and Wi-Fi Load}
\label{tab:headroom}
\centering
\scriptsize
\setlength{\tabcolsep}{3pt}
\begin{tabular}{@{}lrrrrrrrrrr@{}}
\toprule
 & & \multicolumn{5}{c}{\textbf{Polar H10 (pushed)}} & \multicolumn{4}{c}{\textbf{Corsano (pulled)}} \\
\cmidrule(lr){3-7}\cmidrule(l){8-11}
 & \textbf{Load} & \multicolumn{5}{c}{\textbf{\% older than $D$}} & \textbf{Links} & \textbf{Data} & \multicolumn{2}{c}{\textbf{Stale (s)}} \\
\textbf{Hop set} & \textbf{(\%)} & \textbf{100\,ms} & \textbf{250\,ms} & \textbf{1\,s} & \textbf{1\,s$^\ast$} & \textbf{10\,s} & \textbf{lost} & \textbf{(\%)} & \textbf{p50} & \textbf{p99} \\
\midrule
\multicolumn{11}{@{}l}{\emph{Reference}} \\
framing only$^a$ & -  & 80.8 & 54.8 & 0.0  & 0.0  & 0.0 & -  & -  & -  & -  \\
AFH free               & 75  & 89.6 & 62.9 & 0.0  & 0.0  & 0.0 & 0/3 & 99  & 2.0 & 4.0 \\
Busy + 2 clean         & 0   & 89.5 & 62.9 & 0.0  & 0.0  & 0.0 & 0/3 & 99  & 2.0 & 5.0 \\
\addlinespace
\multicolumn{11}{@{}l}{\emph{Forced onto the busy channels, plus $k$ clean}} \\
$k=0^b$                & 75  & 98.8 & 91.9 & 29.3 & 97.7 & 0.0 & 6/6 & 6   & 13  & 76  \\
$k=2$                  & 75  & 95.5 & 75.3 & 0.2  & 1.2  & 0.0 & 0/3 & 51  & 3.9 & 14  \\
$k=4$                  & 75  & 94.6 & 72.0 & 0.0  & 0.2  & 0.0 & 0/3 & 68  & 2.9 & 13  \\
$k=8$                  & 75  & 92.6 & 67.4 & 0.0  & 0.0  & 0.0 & 0/3 & 87  & 2.3 & 7.2 \\
\addlinespace
$k=2^c$                & 100 & 95.0 & 74.1 & 0.0  & 0.2  & 0.0 & 0/3 & 42  & 4.6 & 16  \\
$k=8^c$                & 100 & 92.5 & 67.6 & 0.0  & 0.0  & 0.0 & 0/3 & 89  & 2.5 & 12  \\
\bottomrule
\end{tabular}
\par\smallskip
\parbox{\linewidth}{\scriptsize Busy: data channels 12-19, inside Wi-Fi channel~6. Polar H10: share of samples older than $D$ on arrival, median over runs ($^\ast$worst run). It never lost its link. Corsano: links lost (runs), data received (\% of no load), and staleness, per-run p50 and p99. $^a$Age when the 0.56\,s frame is complete, before any radio delay. $^b$All runs at 75\% load: Polar H10 10 (two sessions, four ran away, two of them up to 1.5\% older than 10\,s), Corsano 6 (session~1). $^c$Polar H10: second session.}
\end{table}

\subsection{Rehabilitation Gym}
\label{sec:results-gym}
\begin{figure}[t]
  \centering
  \includegraphics[width=\columnwidth]{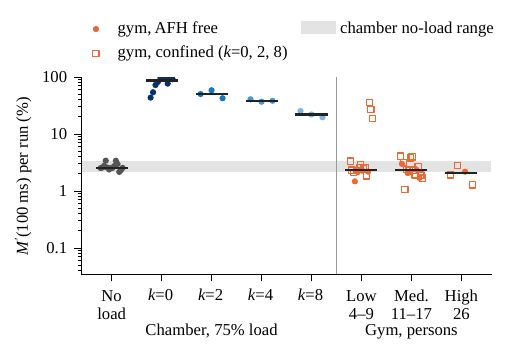}
  \caption{Polar H10 share of samples with added delay over 100\,ms, $M'(100\,\text{ms})$, per run. Chamber: 14 runs without Wi-Fi load and runs at 75\% load by clean channels $k$ (ten at $k=0$). Gym: all 36 runs, by people present at run start. Bars show medians. The three highest gym values correspond to confined runs during the traffic episode.}
  \label{fig:gym-chamber}
\end{figure}

The gym behaved mostly like the quiet chamber (Table~\ref{tab:gym}, Fig.~\ref{fig:gym-chamber}). In 67 of 72 runs, including confined runs without clean channels, both devices stayed close to their chamber no load reference. At most 4.1\% of Polar H10 samples had an added delay over 100\,ms and at most 0.6\% were older than 1\,s, and the Corsano received at least 93\% of its data. The Corsano’s link was stronger in the gym than in the chamber (connection \gls{RSSI} $-61$ to $-70$\,dBm against $-74$\,dBm). The five exceptions were confined runs during or after one traffic episode on the second day, when data traffic on channel~1 rose about 40-fold at 12:08 and 15\,s windows reached 28-70\% airtime. At $k=0$, 4.6\% of Polar H10 samples were older than 1\,s (no growing backlog), and the Corsano disconnected four times and received 6\% of its data, as in the chamber at $k=0$. At $k=2$ and 8, no Polar H10 sample was older than 1\,s, and the Corsano received 58 and 98\% of its data.

\begin{table}[t]
\centering
\caption{Rehabilitation Gym Recordings (No Generated Load)}
\label{tab:gym}
\scriptsize
\setlength{\tabcolsep}{3pt}
\begin{tabular}{@{}>{\raggedright\arraybackslash}p{2.8cm}cccc@{}}
\toprule
 & \multicolumn{2}{c}{\textbf{Day 1}} & \multicolumn{2}{c}{\textbf{Day 2}} \\
\cmidrule(lr){2-3}\cmidrule(l){4-5}
 & \textbf{13:36} & \textbf{16:06} & \textbf{Other} & \textbf{Confined,} \\
 & & & \textbf{runs} & \textbf{from 12:08$^a$} \\
\midrule
People in the gym$^b$        & 9-15 & 5-26 & \multicolumn{2}{c}{4-17} \\
Runs, Polar + Corsano        & 12+12 & 12+12 & 9+9 & 3+3 \\
\addlinespace
\multicolumn{5}{@{}l}{\emph{Wi-Fi, busiest channel}} \\
Channel                      & 1 & 6 & \multicolumn{2}{c}{1} \\
Airtime, median (\%)         & 11.7$^c$ & 6.6 & 12.1 & 21.3$^d$ \\
\quad range (\%)             & 10.2-12.7 & 5.1-14.2 & 10.3-12.7 & 19.0-28.1 \\
Airtime p95, max (\%)        & 13$^c$ & 19 & 14 & 53$^d$ \\
\addlinespace
\multicolumn{5}{@{}l}{\emph{Polar H10}} \\
Older than 250\,ms (\%)      & 63.0 & 63.0 & 62.9 & 68.3 \\
Older than 1\,s (\%)         & 0.0 & 0.0 & 0.0 & 0.0 \\
\quad worst run (\%)         & 0.6 & 0.5 & 0.6 & 4.6 \\
Added delay ${>}$0.1\,s (\%) & 2.3 & 2.3 & 2.2 & 26.8 \\
\addlinespace
\multicolumn{5}{@{}l}{\emph{Corsano}} \\
Links lost (runs)            & 0/12 & 0/12 & 0/9 & 1/3 \\
Data received (\%)           & 99 & 99 & 99 & 58 \\
Stale, p50/p99 (s)$^e$       & 2.0/4.6 & 2.0/4.0 & 2.0/4.0 & 3.5/14 \\
\bottomrule
\end{tabular}
\par\smallskip
\parbox{\linewidth}{\scriptsize Metrics as in Table~\ref{tab:headroom}, medians over runs. $^a$Started during the 12:08-12:29 traffic episode or after the monitor stalled. The Corsano $k=0$ run lost its link four times, received 6\% of its data, and was stale for 13/79\,s (p50/p99). $^b$Hand counts about every 30\,min. $^c$Three runs, before the monitor fault at 14:04. $^d$Four runs, as the monitor stalled at 12:29. $^e$In two runs per Day-1 session, the watch wrote no new bio-impedance data for about 5\,min while transfers continued. Session details in Supplementary Section~\ref{S-sup:gym}.}
\end{table}

\subsection{End to End through MEDWEAR}
\label{sec:results-e2e}
MEDWEAR committed every replayed message, with no duplicates or checksum mismatches, except \gls{ecg} batches rejected by the signal plausibility check: one batch of 146 samples in the 120 chamber recordings and 0.27\% of \gls{ecg} samples in the gym (Supplementary Section~\ref{S-sup:replay}). On the live path, which runs the ROS4HC driver and \gls{ros2} before any MEDWEAR component, a \gls{ros2} subscriber received each frame 3\,ms (median) and at most 7.5\,ms (99th percentile) after the host, in every condition of Table~\ref{tab:e2e} including $k=0$. The durable path ($W=0.1$\,s) added 0.14\,s (median), mostly batching, as the \texttt{ros2_to_medwear} bridge held each sample 122\,ms on average and queued its batch for 0.3\,ms. That suits alarms but not control loops: at commit, 88-95\% of samples were older than 250\,ms, whereas 1 and 10\,s deadlines were missed for at most 1.0 and 0.1\% of samples in the median run outside $k=0$ (38 and 0.3\% at $k=0$, as samples were still queued in the strap). Larger windows trade freshness for load: $W=5$\,s cut requests from 1.54 to 0.18/s but raised the mean batch wait from 0.12 to 2.7\,s.

\begin{table}[t]
\caption{Polar H10 ECG Live-Path Latency and Deadline Misses at the Host and on the Durable Path ($W=0.1$\,s)}
\label{tab:e2e}
\centering
\scriptsize
\setlength{\tabcolsep}{3pt}
\begin{tabular}{lrrrrr}
\toprule
\textbf{Condition} & \textbf{Live latency} & \multicolumn{3}{c}{\textbf{Host: age $>D$ (\%)}} & \textbf{Durable$^\ast$} \\
\cmidrule(lr){3-5}
 & \textbf{p50/p99 (ms)} & \textbf{100\,ms} & \textbf{250\,ms} & \textbf{1\,s} & \textbf{$M$(10\,s)} \\
\midrule
AFH free, 100\% load & 3.0/6.3   & 89.7 & 63.0 & 0.0 & 0.1 \\
$k=8$, 75\% load     & 3.0/6.8   & 92.6 & 67.4 & 0.0 & 0.1 \\
$k=4$, 75\% load     & 2.9/6.3   & 94.6 & 72.0 & 0.0 & 0.1 \\
$k=2$, 75\% load     & 3.0/7.5   & 95.5 & 75.3 & 0.2 & 0.1 \\
$k=0$, 75\% (10 runs) & 3.0/6.7 & 98.8 & 91.9 & 29.3 & 0.3 \\
\bottomrule
\end{tabular}
\par\smallskip
\parbox{\linewidth}{\scriptsize Live latency: $t_\text{r}-t_\text{h}$ on the original recordings. Durable: replay. Three runs per condition, at $k=0$ all ten runs of Table~\ref{tab:headroom} (five $k=0$-labelled and five confined, four of them runaways). Host: sample age on arrival, medians, as in Table~\ref{tab:headroom}. $^\ast$Durable: sample age at commit, median of runs, over the full window, including samples queued at the end. Host age over the same window gives the same values except in three of the four runaways (2.7-3.9 vs.\ 2.8-4.4\%).}
\end{table}

\subsection{Ingestion of Buffered Data}
\label{sec:results-scale}
For devices that buffer data, such as the Corsano, what the live path misses can still be collected later. From the CARDAMON study, MEDWEAR stored 87,041 \gls{ecg} batches (June to August 2026) under 32 participant identifiers, including test entries. Across all uploads, validation rejected 31.2\% of requests on first submission, all for timestamp misalignment in the vendor exports, which, stored as received, would have carried wrong times. After correction and resubmission, the idempotency key rejected all duplicate submissions.

Uploading later is safe only if acknowledged data are stored and a retry cannot store them twice. In single-node load and fault tests (Supplementary Section~\ref{S-sup:benchmark}), MEDWEAR stored a resubmitted batch only once and lost no acknowledged sample in 90 container faults. Requests cut off between its two writes left their batches pending until a retry completed them. With the per-user rate limit raised, since all simulated devices shared one identity, its single \gls{API} worker served up to 50 batches/s, but only 38\% and 17\% of arrivals completed at 100 and 200 devices (Supplementary Fig.~\ref{S-fig:s-load}). MEDWEAR stored 5 of 15 malformed requests, currently allowed: a body with a wrong content type is parsed as \gls{JSON}, a missing sampling rate is derived from the timestamps, a missing or unknown project falls back to the user's own, and the schema identifier is a free-text version label (samples are validated by signal type).

\section{Implementing a Real-Time Wearable System with MEDWEAR}
\label{sec:guideline}
Starting from the consumer's deadline $D$ and the $M(D)$ it tolerates (Table~\ref{tab:deadlines} can be used as a reference), we propose:
\begin{enumerate}
  \item Write the driver with ROS4HC \cite{manriquez-cisterna_ros_2026}, publishing \texttt{healthcare\_msgs} \cite{perez_serrano_healthcare_msgs} (including the sensor's timestamps) or the standardized \gls{JSON} schemas.
  \item Characterize the sensor in the intended environment with MEDWEAR's profiling tool, at quiet and busy times. Be aware that three runs per condition screen for gross failures but cannot rule out rare runaways. 
  \item Check whether the sensor can match the deadline and iterate on the design if needed, e.g. change the sensor or what it streams, move Wi-Fi load to 5\,GHz.
  \item Monitor sample age during operation and watch the gateway's \gls{AFH} channel map. The system shall fall back to a safe state when sample age exceeds $D$.
\end{enumerate}

\section{Discussion}
\label{sec:discussion}
Real-time consumers of Healthcare~5.0 need their data within a deadline to function as intended, so they need to know the age of each sample.

However, each setup is unique: sensors, \gls{BLE} controllers, firmware, and environments differ. Generalizing to every scenario is therefore outside the scope of this work. Instead, we show that data can arrive late while the link stays connected, and we release MEDWEAR's profiling tool \cite{medwear_software_2026} so that each integrator can measure sample age on their own configuration.

The gym recordings agreed with hospital surveys \cite{al_kalaa_long_2016,al_kalaa_characterizing_2017}: the 2.4\,GHz band is lightly used, with occasional bursts. This finding is interesting for two reasons. First, there is generally enough headroom to introduce the use of wearable sensors in such spaces. For the Polar H10, the link occupied the central’s radio for 1.25\% of the time (22.2 connection events per second, 0.56\,ms each on average, with \gls{AFH} free). So one \gls{BLE} controller would theoretically have radio time for about 80 Polar H10 devices. This is only an upper reference value, since the controller's connection limit would likely bound it first, and links on other controllers would add collisions that \gls{AFH} cannot avoid \cite{la_dense_2018}.

Second, systems should be designed to handle occasional bursts. On the sender and receiver side, previous work has handled such conditions with an adaptive \gls{MAC} protocol \cite{hassan_adaptive_2024}, informed hopping \cite{poirot_eafh_2022}, real-time scheduling \cite{li_rtble_2023}, a connection interval adapted at runtime \cite{spork_improving_2020}, or by changing what the sensor sends \cite{tipparaju_mitigation_2021}. These solutions require modifying either the wearable or the gateway's \gls{BLE} controller, which commercial devices generally do not allow.

Until commercial devices allow such changes, an integrator can control at most the host side, and monitor sample age, so that downstream algorithms can discard late samples or fall back to a safe state.

\subsection{Limitations}
\label{sec:limitations}
The study uses one unit of each device, one controller (Intel AX211, BlueZ), one access point on one channel, one distance, and sensors on a sponge, so body shadowing and motion are not modeled. On a body, the Polar H10's link could weaken toward the Corsano's $-74$\,dBm, so our setup is likely optimistic. The Corsano also differed in signal strength, transfer model, firmware, driver, and encryption. Overlap was set through the host channel map, and the interferer was never local to the peripheral, a case that can defeat \gls{AFH} even with clean channels \cite{spork_improving_2020}. Ages are best-case lower bounds without a constant radio-and-host floor, which affects absolute values near 100\,ms but not comparisons between conditions or deadlines of 1\,s and longer.

\section{Conclusion}
\label{sec:conclusion}
This work characterized sample age from acquisition to storage in MEDWEAR, using a commercial \gls{ecg} strap (Polar H10), in an anechoic chamber and a rehabilitation gym, and compared it against illustrative clinical deadlines.

The main source of delay depended on radio conditions. In the chamber, with free \gls{AFH} hopping, the strap's batching was the main contributor, as it collects 73 samples before sending one packet, making samples on average 280\,ms and up to 550\,ms old before transmission, so 81\% already exceeded the 100\,ms deadline. The rest of the chain added little: the \gls{BLE} link about 45\,ms (median added delay above the run's minimum, about one connection interval), the \gls{ros2} live path 3\,ms, and MEDWEAR's durable path 140\,ms (mostly from batching data).

When the link was forced onto the loaded channel, the link itself became the main source of delay instead. The strap remained connected, but accumulated backlogs of up to 11\,s, with a median of 29\% and up to 98\% of samples missing the 1\,s deadline, whereas the Corsano disconnected. With at least two clean channels, the strap's miss rate was at most 1.2\%.

The gym generally behaved like a quiet chamber, except for one traffic burst, which made 4.6\% of samples older than 1\,s in runs confined to the busy channel. This traffic behavior is in line with hospital surveys \cite{al_kalaa_long_2016,al_kalaa_characterizing_2017}.

These results show that radio congestion and sensor-side batching can compromise data freshness. Wearables for real-time applications should, therefore, be profiled in their intended environments and their sample age should be monitored. MEDWEAR \cite{medwear_software_2026}, the wireless experiments \cite{wireless_experiments_software_2026}, and the chamber recordings \cite{perez_serrano_2026_23750299} are openly available.

\section*{Acknowledgment}
The authors thank Prof.\ Christoph Studer and Alexander Stutz-Tirri of the Integrated Systems Laboratory (IIS), ETH Z\"urich, for access to and support with their anechoic chamber, the Swiss Paraplegic Center (SPZ), Nottwil, and its therapy team for allowing recordings in the rehabilitation gym during regular operation, and the CARDAMON study participants.
The authors used Claude (Anthropic) to assist with code development, improve figure presentation, and improve the language and readability of the manuscript. All AI-assisted code, figures, and text were reviewed, verified, and edited by the authors, who take full responsibility for this article.

\section*{References}

\bibliography{refs_ieee}

\clearpage
\onecolumn
\includepdf[pages=-,pagecommand={\thispagestyle{empty}}]{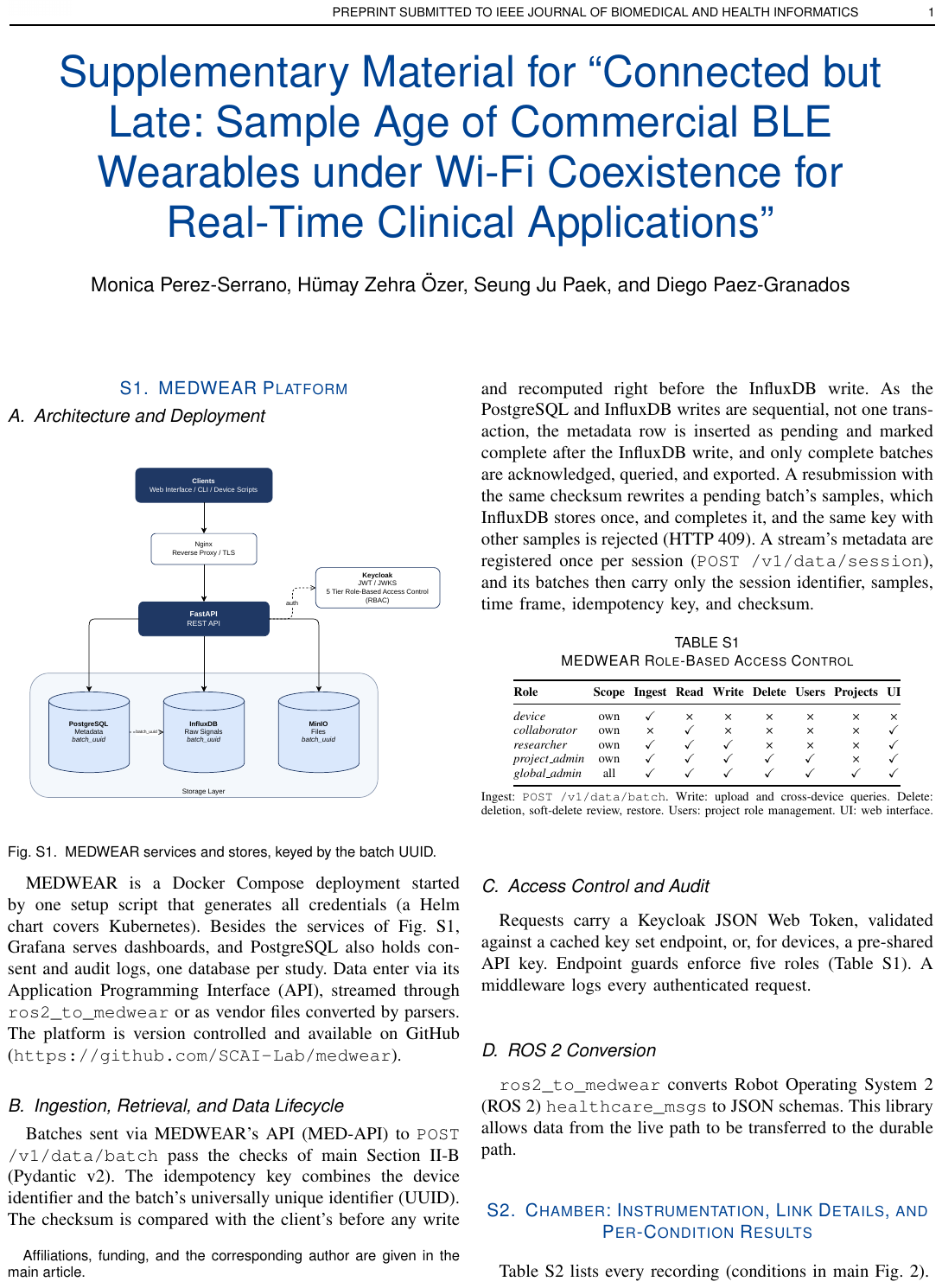}

\end{document}